\documentclass[aps, pra, twocolumn, showpacs, byrevtex]{revtex4-1}
 \usepackage[english]{babel}
 \usepackage{amssymb,amsmath}
 \usepackage{color}
 \usepackage{pstricks}
 \usepackage{bbold}
 \usepackage{graphicx}
 \usepackage{epsfig}
 \usepackage{color}
 \usepackage{verbatim}

\begin{document}

\title{Pairing of few Fermi atoms in one dimension}

\author{Pino D'Amico}
\email{pino.damico@nano.cnr.it}

\affiliation{CNR-NANO Research Center S3, Via Campi 213/a, 41125 Modena, Italy}

\author{Massimo Rontani}
\email{massimo.rontani@nano.cnr.it}

\affiliation{CNR-NANO Research Center S3, Via Campi 213/a, 41125 Modena, Italy}

\begin{abstract}
We study a few Fermi atoms interacting through attractive contact forces in
a one-dimensional trap by means of numerical exact diagonalization.
From the combined analysis of energies and
wave functions of correlated ground and excited states
we find evidence of BCS-like pairing even for very few
atoms. For moderate interaction strength, we reproduce the even-odd 
oscillation of the separation energy observed in 
[G. Z{\"u}rn, A. N. Wenz, S. Murmann, 
A. Bergschneider, T. Lompe, and S. Jochim, 
Phys. Rev. Lett. {\bf 111,} 175302 (2013)]. 
For strong interatomic attraction  
the arrangement of dimers 
in the trap differs from the homogeneous case 
as a consequence of Pauli blockade
in real space.
\end{abstract}


\maketitle

\section{Introduction}

Pairing between fermions is a basic phenomenon emerging
in quantum degenerate systems as diverse as electrons in
metals \cite{deGennes1999}, protons
and neutrons in nuclei \cite{Migdal1960,Bohr2008}
and neutron stars \cite{Ginzburg1964,Pines1969},
$^3$He atoms \cite{Legget2006},
electrons and holes in semiconductors \cite{Rontani2014},
cold atoms confined in magneto-optical 
traps \cite{Regal2003,Bartenstein2004,Zwierlein2005,Legget2006,Bloch2008,Giorgini2008}.
In nuclei, pairing enhances the stability of isotopes  
with an even number of constituents, reaching the maximum
at the closure of an energy shell 
\cite{Bohr1958,Migdal1960,Bohr2008,Zelevinsky2003,Brink2005}.
In metals, electrons of opposite spins form
Cooper pairs that condense in the superconducting phase, as 
explained by the weak-coupling theory 
by Bardeen, Cooper and Schrieffer (BCS) \cite{BCS1957}.

Experiments with cold Fermi atoms provide unprecedented control on both
the shape of the trap confinement potential and
the interatomic interaction 
strength---the latter by sweeping a magnetic offset field 
through a Feshbach resonance 
\cite{Chin2010}.
This enables novel possibilities, like to explore the transition from BCS-like
superfluidity to Bose-Einstein condensation (BEC)
of strongly bound atom dimers 
\cite{Regal2003,Legget1980,Nozieres1985,Bartenstein2004,Zwierlein2005},
to control the atom number $N$ with unit precision---down to the empty-trap
limit \cite{Serwane2011,Zuern2012,Zuern2013,Wenz2013}, 
as well as to change the dimensionality of the 
system \cite{Bloch2008,Moritz2005,Liao2010,Martiyanov2010,Feld2011,Dyke2011,Sommer2012}.

In these tunable traps, the pairing gap $\Delta$---the order 
parameter of
the superfluid phase---may deviate from the expectations
for homogeneous systems and exhibit a significant dependence 
on the atom number
$N$ as well as on the dimensionality $d$ 
\cite{Heiselberg2002,Bruun2002,Heiselberg2003}.
This is seen from BCS gap equation,
which allows for a finite value of $\Delta$ provided
the density of states $g(\varepsilon)$
is large at the Fermi surface $\varepsilon_F$
($\varepsilon$ is the energy reckoned from
the bottom of the trap).
This density is enhanced by the occurrence
of energy shell degeneracies at higher dimensions,
depending on $d$ as
$g(\varepsilon)\sim\varepsilon^{d-1}$ on a coarse-grain energy 
scale \cite{Pethick2002}.
Therefore, pairing is harder to accomplish
at lower dimensions as it requires stronger inter-species 
attraction. Besides, the spiked features of 
$g(\varepsilon)$ on the fine energy scale 
might make $\Delta$ strongly fluctuate when filling successive shells.

\begin{figure}[htbp]
\centering
\includegraphics[width=80mm, angle=0]{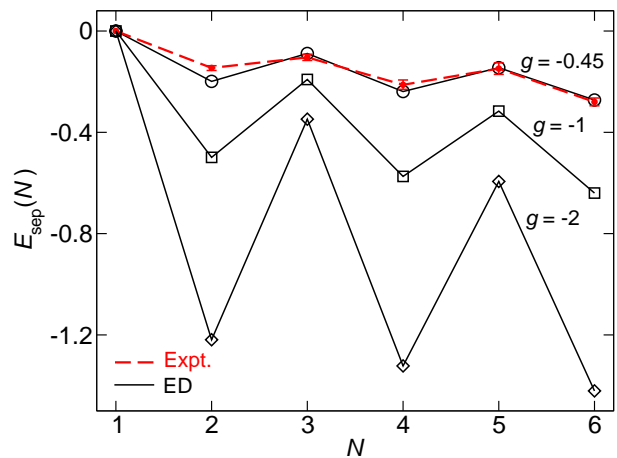}
\caption{(color online)
Separation energy $E_{\text{sep}}(N)$ vs atom number 
$N$. Black circles, squares, diamonds with solid lines correspond to 
$g=$ -0.45, -1, -2, respectively. Red [gray] circles with error bars
and dashed lines
are the measured data reported in \cite{Zuern2013}. 
The energy unit is $\hbar\omega$. 
Lines are guides to the eye.}
\label{separation_energy_fig}
\end{figure}
A recent experiment by the Heidelberg group seems at odds with these 
expectations \cite{Zuern2013}. The magneto-optical trap was effectively 
one-dimensional (1D) as the aspect ratio was 1:10, 
the Fermi energy $\varepsilon_F$ was
comparable to the longitudinal oscillator spacing, and the temperature
was around half the Fermi 
temperature \cite{Serwane2011}.
The trapping potential was deformed to 
measure the time spent by $^6$Li atoms to
tunnel out of the trap. This decay time was then linked to
the separation energy of the system with $N$ 
fermions \cite{Rontani2012,Rontani2013b}, which exhibited 
a regular even-odd oscillation vs $N$ for moderate attraction 
strength
and very small atom number, $N\le 6$,
as shown in 
Fig.~\ref{separation_energy_fig} (dots with error bars and dashed lines). 
This alternate staggering was attributed
to pairing, in analogy with similar data
for neutron separation experiments in nuclei \cite{Bohr2008}. 
Intriguingly, BCS theory predicts that the pairing gap $\Delta$ vanishes 
exponentially with the interaction (see violet [gray] 
curve in Fig.~\ref{pairing_gap}).

These findings call for a theoretical analysis beyond
mean-field BCS level, to take into account both finite-size 
fluctuations \cite{vonDelft2001,Viverit2004,Olofsson2008,Forbes2011,Ribeiro2012,Angelone2014}
and correlations at all orders.
The problem of 1D Fermi gas with short-range
interactions was solved exactly only for the 
homogeneous system, through either Bethe ansatz 
\cite{Gaudin1967,Yang1967,Takahashi1971,Guan2013} 
(known as Gaudin-Yang model for attractive interactions) 
or mapping to the Luttinger 
Hamiltonian \cite{Yang2001,Giamarchi2003,Cazalilla2011}.
Therefore, available results
\cite{Astrakharchik2004b,Tokatly2004,Fuchs2004,Guan2007,Guan2009,Guan2013} are  
useless for the harmonic trap analyzed here.
On the other hand,
the small number of $^6$Li atoms studied in \cite{Zuern2013} allows for 
comparison with numerical exact diagonalization (ED), which 
provides energies and wave functions of both ground and excited states 
\cite{Rontani2006,Kalliakos2008,Rontani2009b,Singha2010,Wang2012b,Pecker2013,Sowinski2013,DAmico2014}, whereas quantum 
Monte Carlo simulations are restricted to the ground state \cite{Juillet2004,Giorgini2008,vonStecher2008,Casula2008,Zinner2009,Gilbreth2013}.

In this paper we investigate theoretically the pairing between 
a few Fermi atoms populating a 1D harmonic trap.
From the analysis of both ground- and excited-state ED energies
we find that the pairing gap $\Delta$ is well defined even at small $N$,
recovering the measured even-odd effect
(black circles and solid lines in Fig.~\ref{separation_energy_fig}).
The ED wave function is significantly affected by interaction already
at moderate coupling strength, close to the regime  
achieved in the experiment \cite{Zuern2013}.
For strongly bound dimers, the pair wave function  
exhibits a peculiar $N$-dependent spatial modulation 
that is absent in the bulk. This unexpected behavior---a manifestation of
Pauli blockade in real space---may be observed  using 
time-of-flight techniques.

The structure of this paper is as follows.
We introduce the system Hamiltonian and the ED method 
in Sec.~\ref{s:model}. 
Then we compare the ED
separation energy with the measured data, also in connection   
with the fundamental energy gap (Sec.~\ref{s:separation}).
We evaluate the pairing gap $\Delta$ in two complementary ways,
considering both ground-state energies by changing $N$ 
and excited-state energies for fixed $N$ (Sec.~\ref{s:pairing}).  
We access the correlated ground state by computing the
pair correlation function $G(x)$, which allows us to estimate
the size of Cooper pairs (Sec.~\ref{s:pair}).
We eventually focus on the BEC-like regime of strong attraction,
showing that $G(x)$ departs from the bulk behaviour due to Pauli
blockade in real space (Sec.~\ref{s:Pauli}).
After Conclusions, Appendix \ref{a:Gbulk} illustrates the 
derivation of the bulk pair
correlation function $G(x)$ plotted in Fig.~\ref{Gx}(f).

\section{Exact diagonalization}\label{s:model}

We consider $N$ atoms of spin $1/2$ 
confined in a 1D harmonic trap of frequency $\omega$
and
interacting through an attractive contact force,
\begin{equation} 
 H = \sum_{i=1}^{N}\left[\frac{p^2_i}{2m} 
+ \frac{1}{2}m\omega^2 x^2_i\right]+g'\sum_{i<j}\delta(x_i-x_j),
\label{eq:H}
\end{equation}
where $g'<0$ is the coupling constant and
$m$
is the atom mass.
Throughout this article we
use $\hbar\omega$ as energy unit and 
$\ell = (\hbar/ m \omega)^{1/2}$ as length unit, hence
the dimensionless coupling constant is $g=g'/(\hbar\omega\ell)$.

The ED wave function is the superposition of those Slater determinants
obtained by filling the lowest $N_{\text{orb}}$ harmonic-oscillator orbitals
with $N$ fermions in all possible ways 
(also known as full configuration interaction \cite{Rontani2006}).
In this Fock space the Hamiltonian \eqref{eq:H} is a sparse matrix,
with blocks labeled by the total spin projection $S_z$, 
parity, and $N$.  
The maximum linear size of the eigenvalue
problem (for $N=6$ and $N_{\text{orb}}=25$) is 2,644,928, 
which we solve with the home-built
parallel code DONRODRIGO \cite{Rontani2006}. 
The ED convergence is demanding 
in the present attractive regime, as the method just mimics
the cusp of the exact wave function induced by
the contact interaction \cite{DAmico2014}. 
Therefore, the choice of $N_{\text{orb}}$ is 
the trade-off between 
accuracy and computational load, the Fock space size scaling 
exponentially with $N$. 
Here we used $N_{\text{orb}}=25$,
with an error on the ground state absolute interaction
energy of 5.6, 17, 22 \% for $g=-1$, -2, -3, respectively, 
and $N_{\text{orb}}=50$ in  
the illustrative case $N=3$ and $g=-4$ of Fig.~\ref{Gx},
which gives an error of 18 \%. 
The accuracy on $\Delta$ is much higher due 
to mutual cancellation of systematic errors.   

\section{Separation energy}\label{s:separation}

The key quantity we obtain from ED ground-state energies $E_0(N)$
is the chemical potential 
\begin{equation}
\Delta_1(N)=E_0(N)-E_0(N-1).
\end{equation}
This is the lowest resonating energy
of the $N$th atom tunneling out of the trap while leaving the other 
$N-1$ atoms in the trap 
in their ground state \cite{Rontani2009b,Rontani2012,Rontani2013b}. 
The separation energy $E_{\text{sep}}(N)$ is the net interaction energy
contributing to $\Delta_1$,
\begin{equation}
E_{\text{sep}}(N) = \Delta_1(N) - \Delta_1^*(N),
\end{equation}
with $\Delta_1^*$ being 
the chemical potential in the absence of interaction.
The magnitude of $E_{\text{sep}}$ 
is the contribution to the ionization energy 
due to interatomic attraction.

Figure \ref{separation_energy_fig} shows even-odd 
oscillations of $E_{\text{sep}}$ as the trap is filled with atoms.
The ED spectrum (black circles with solid lines) fits well the
measured data of \cite{Zuern2013} (red [gray] circles with error bars
and dashed lines)
for $g=-0.45$. This value reasonably compares with
the experimental estimate of $g\sim -0.9$
(in our units), as the 
trap was strongly deformed with respect to the harmonic potential
to allow the escape of atoms \cite{note_g09}. A possible reason for the 
residual mismatch between theory and experiment is the 
anharmonicity of the actual energy spacing in the trap.

In Fig.~\ref{separation_energy_fig} both peak-to-valley ratios and 
magnitudes of absolute minima increase with attraction strength.
Besides, the minima are deeper at higher atom numbers.
These features are consistent with a
BCS-like scenario,
since: (i) the BCS ground state is more stable for even $N$, as
all atoms are paired (ii) its energy
gain increases with $N$---a signature of collective effect. 

\begin{figure}[htbp]
\centering
\includegraphics[width=80mm, angle=0]{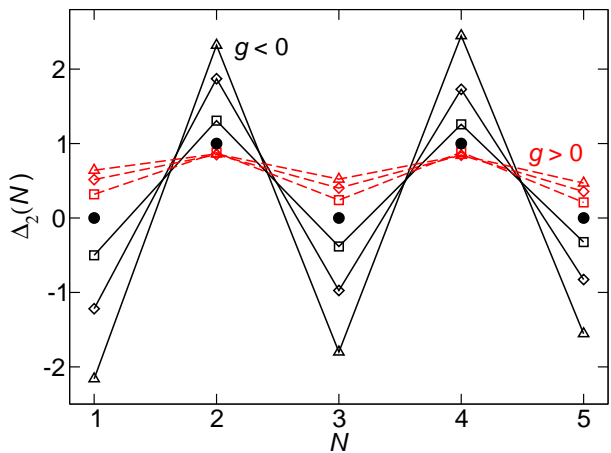}
\caption{(color online)  
Fundamental energy gap $\Delta_2(N)$ vs atom number $N$. 
The black (red [gray]) color with solid (dashed) lines 
points to attractive (repulsive) interaction.
Squares, diamonds, triangles correspond to $|g|=1, 2, 3$ respectively. 
The filled black circles 
are noninteracting data ($g=0$). 
The energy unit is
$\hbar\omega$.
Lines are guides to the eye.}
\label{fundamental.gap_fig}
\end{figure}
However, the observed even-odd oscillation of $E_{\text{sep}}$
might have a different explanation, being simply 
due to the filling of successive twofold degenerate trap orbitals.
To clarify this matter we plot in Fig.~\ref{fundamental.gap_fig}  
the fundamental energy gap 
\begin{equation}
\Delta_2(N)=\Delta_1(N+1) -\Delta_1(N), 
\end{equation}
which is the difference 
between the tunneling energies of the atom added to and removed from 
the trap \cite{Rontani2009b}.
Here, it is instructive to consider
repulsive (red [gray] symbols with dashed lines) interactions
as well as attractive forces (black symbols with solid lines),
since in both cases $\Delta_2$ exhibits an even-odd oscillation.
At small coupling ($\left|g\right|=1$, square symbols) both patterns 
slightly deviate from the staggering noninteracting 
sequence (filled black circles), 
hence $\Delta_2\approx 1$ for even $N$ and $\approx 0$ for odd $N$,
the energy separation between consecutive orbital levels being unity.
As $\left|g\right|$ increases $\Delta_2$ changes qualitatively 
depending on the interaction sign.
Strong repulsive interactions wash out the staggering of $\Delta_2$, which
tends to a homogeneous positive value \cite{Wang2012b}.
On the contrary, strong attractive forces enhance 
even-odd oscillations, suggesting BCS-like pairing.
Indeed,
if $N$ is even, all atoms form singlet pairs and a large amount of positive
energy $\Delta_2$ is required to add one unpaired atom.
For odd $N$, the fundamental gap $\Delta_2(N)$ is large and \emph{negative},
since energy is gained by pairing with an opposite-spin atom.

\begin{figure}[htbp]
\setlength{\unitlength}{1 cm}
\begin{picture}(8.5,5.5)
\put(0.2,-0.2){\epsfig{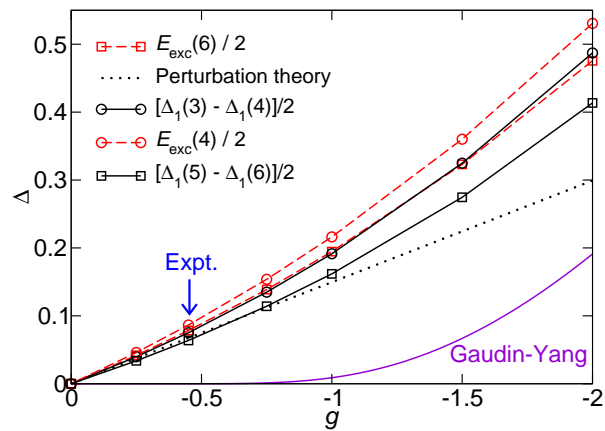}}
\end{picture}
\caption{(color online)
Complementary estimates of the pairing gap $\Delta$ vs 
interaction strength $g$.  
The violet [gray] curve is the exact prediction of
Gaudin-Yang model in the BCS limit with 
$\varepsilon_F=3/2=\Delta_1^*(N=4)$.
The dotted line is the prediction by perturbation theory whereas
the remaining solid and dashed lines are guides to the eye.
We use $\hbar\omega$ as energy unit and 
$\ell = (\hbar/ m \omega)^{1/2}$ as length unit, hence
the dimensionless coupling constant is $g=g'/(\hbar\omega\ell)$.
}
\label{pairing_gap}
\end{figure}

\section{Pairing gap}\label{s:pairing}

The computation of $\Delta_1$ allows us
to evaluate the pairing gap $\Delta$ 
from two consecutive chemical potentials,
\begin{equation}
\Delta = \frac{ \Delta_1(N)-\Delta_1(N+1) }{2},
\label{eq:Delta_stagger}
\end{equation}
with $N$ odd \cite{Giorgini2008,Rontani2009b}.
Here the sign change is due to the staggering of 
$\Delta_1$, which alternately points to an energy expense and gain
respectively for adding an unpaired atom and matching all pairs.
Reassuringly, the pairing gaps $\Delta$ obtained for $N=3$
(black circles and solid lines in Fig.~\ref{pairing_gap}) 
and $N=5$ (black squares and solid lines)
exhibit a similar dependence
on $g$, coinciding within 15\% at worse at large interaction strength
$g=-2$. We see that $\Delta$ is a convex 
function of $g$, smoothly rising up to the value $\Delta\sim 0.5$, which is
of the order of level spacing. 
The behavior of $\Delta$ is 
similar to that predicted in 2D \cite{Rontani2009b}
and 3D \cite{Zinner2009} for few atoms, suggesting
that intra- and intershell contributions to pairing 
\cite{Heiselberg2002,Bruun2002,Heiselberg2003} are comparable.

\begin{figure}[htbp]
\setlength{\unitlength}{1 cm}
\begin{picture}(8.5,6.5)
\put(0.2,-0.2){\epsfig{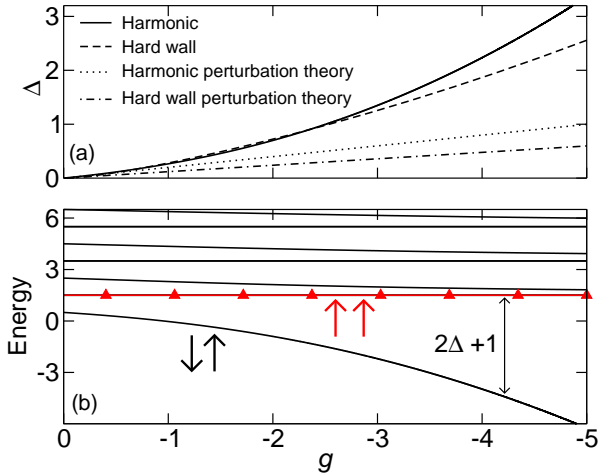}}
\end{picture}
\caption{(color online)
Extraction of the pairing gap $\Delta$ from the excitation 
spectrum of two atoms.
(a) Paring gap $\Delta$ vs interaction strength $g$. 
The solid (dashed) curve refers to the exact result for two 
fermions with parallel spins in a harmonic (hard-wall) trap,
with the harmonic oscillator length $\ell=(\hbar/m\omega)^{1/2}$
being equal to the width of the hard-wall quantum well.
Dotted and dashed-dotted lines are the predictions of perturbation
theory at first order in $g$ for
the harmonic and hard-wall traps, respectively.
(b) Low-lying energy spectrum of two-fermions in the harmonic trap
in the relative frame vs $g$. 
$\Delta$ is inferred from the spin excitation gap
separating the two lowest energy branches,
which are respectively the lowest black line
(balanced system with $S_z = 0$)
and the red (gray) line with triangles
(unbalanced system with  $S_z = \pm 1$).
The energy unit is $\hbar\omega$  
and
the dimensionless coupling constant is $g=g'/(\hbar\omega\ell)$.
}
\label{figS1}
\end{figure}

A complementary study of the paring gap $\Delta$ relies on the
ED excitation spectrum for fixed $N$. For the sake of clarity,
we first focus on the paradigmatic case $N=2$, whose exact solution
is known analytically \cite{Busch1998,DAmico2014}.
The low-lying energy spectrum for relative motion
is shown in Fig.~\ref{figS1}(b),
limitedly to negative
interaction strength $g$. There are two distinct families of
energies branches, differing in orbital parity.
The lines that vary with $g$ correspond to states that 
are even under particle exchange and hence associated with atoms 
of opposite spin, with $S_z=0$. The horizontal lines, independent from $g$, 
are the energies of two
atoms of like spins  with $S_z=\pm 1$, 
whose contact interaction is void as the 
orbital wave function is odd.
Note that, in this odd sector, energy levels are degenerate 
with multiple center-of-mass excitations.

We link the paring gap $\Delta$
to the spin excitation gap \cite{Fuchs2004,Giorgini2008},
which here is the energy difference between the two lowest 
energy branches, highlighted in Fig.~\ref{figS1}(b).
Clearly, we require $\Delta$ to vanish
in the noninteracting limit $g\rightarrow 0$. Therefore, 
we subtract from the excitation gap
a residual excitation energy quantum, which
is unrelated to interactions and absent in the bulk.
The remaining excitation gap is expected to be twice the
gap $\Delta$ for BCS-like pairing, 
since the spin flip leaves 
two atoms unpaired \cite{Brink2005,Giorgini2008,Rontani2009b}.

\begin{figure}[htbp]
\setlength{\unitlength}{1 cm}
\begin{picture}(8.5,6.2)
\put(0.2,-0.2){\epsfig{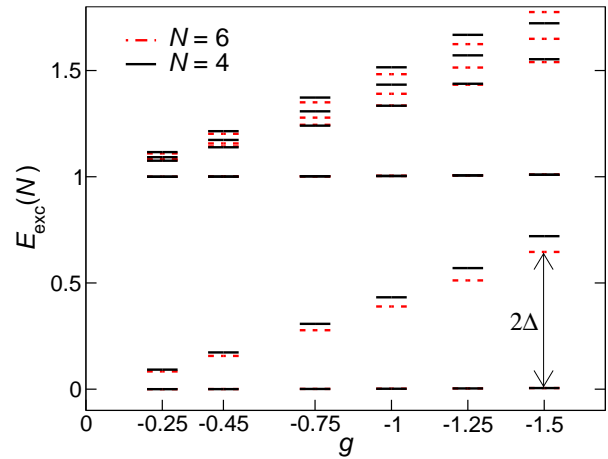}}
\end{picture}
\caption{(color online)  
Excitation energies $E_{\text{exc}}(N)$
for $N=4$ (black lines) and $N=6$ 
(red [gray] dashed lines) vs $g$ for balanced spin population ($S_z=0$). 
Energies were referenced to the ground state after subtracting 
the first center-of-mass excitation quantum. The two lowest branches
are degenerate with those of the unbalanced system with 
$S_z = \pm 1$.
We use $\hbar\omega$ as energy unit and 
$\ell = (\hbar/ m \omega)^{1/2}$ as length unit, hence
the dimensionless coupling constant is $g=g'/(\hbar\omega\ell)$.
}
\label{exc}
\end{figure}

The extracted value of $\Delta$ for $N=2$,
shown in Fig.~\ref{figS1}(a) as a solid line, compares well with similar
data obtained for higher atom numbers, as illustrated
in Figs.~\ref{pairing_gap} and \ref{exc}.
In Fig.~\ref{exc} we plot
the lowest excitation energies $E_{\text{exc}}(N)$
of the system with
$N = 4$ (black lines) and $N=6$ (red [gray] dashed lines) and
$S_z=0$, i.e., balanced spin population.
Here we have referenced all energies to 
the ground state after subtracting the first center-of-mass excitation 
quantum. Again, multiple center-of-mass excitations are unrelated
to atom-atom correlations and hence independent from $g$, 
as shown in Fig.~\ref{exc} for the second excitation quantum.
As the interaction strength $\left|g\right|$
increases an energy gap develops
generically, since pairs must be broken to excite the system.
We find that, due to the symmetry of Hamiltonian \eqref{eq:H}, 
the two lowest excitations 
shown in Fig.~\ref{exc} are degenerate with those obtained 
by flipping one atom spin
($S_z= \pm 1$), as they are connected by a rotation in spin space. 
Therefore, we take these excitations to be twice the gap $\Delta$  
(label in Fig.~\ref{exc}). 

The estimate of $\Delta$ extracted
from the excitation spectrum of Fig.~\ref{exc} is plotted
in Fig.~\ref{pairing_gap} 
for $N=6$ (red [gray] squares and dashed lines) and
$N=4$ (red [gray] circles and dashed lines). 
The good overall matching between
these excitation gaps and the staggering-energy gaps
discussed before (black symbols and solid lines) shows that a 
BCS-like pairing gap $\Delta$ 
emerges already for very few fermions, being relatively insensitive
to finite-size fluctuations.

However, the magnitude of $\Delta$ significantly exceeds 
the BCS bulk value 
\begin{equation}
\Delta = \frac{8}{\pi} \varepsilon_F \sqrt{\frac{\left|\gamma\right|}
{\pi}}
\exp{\left(-\frac{\pi^2}{2\left|\gamma\right|} \right)} 
\label{eq:Delta_BCS}
\end{equation} 
(in standard units),
which is shown by the
violet (gray) curve in Fig.~\ref{pairing_gap}, being
the exact solution of Gaudin-Yang model 
in the limit $\gamma \rightarrow 0-$ 
\cite{Gaudin1967,Fuchs2004}. 
This discrepacy hardly depends on the Fermi energy $\varepsilon_F$
that enters 
the 
coupling constant
$\gamma=(g\pi/\hbar)(m/8\varepsilon_F)^{1/2}$ 
[here $\varepsilon_F=3/2=\Delta_1^*(N=4)$],
since in the bulk $\Delta$ 
vanishes exponentially---a non-perturbative result---whereas
in the trap
$\Delta$ scales almost linearly with $g$ 
up to
$g\approx -1$. 

This latter trend is 
well reproduced by perturbation theory, as shown by 
the dotted line 
in Fig.~\ref{pairing_gap}, using the
estimate $\Delta = -3g/ (8\sqrt{2\pi})$, which is obtained
by first averaging the interaction over the noninteracting ground states
and then using these energy corrections for the staggering-energy 
definition \eqref{eq:Delta_stagger}
of $\Delta$ with $N=3$.
Therefore, for
the experiment \cite{Zuern2013}
(arrow in Fig.~\ref{pairing_gap}),
the wave function
is substantially unaffected by interatomic correlations.
For stronger interactions $\Delta$ significantly deviates from linearity,
as a consequence of
co-operative effects.

We attribute the departure of the functional
form of $\Delta$ from the bulk exponential behaviour \eqref{eq:Delta_BCS}
to a genuine manifestation of few-body physics. In fact,
whereas perturbation theory converges
for the finite system, 
at least for small values of $g$,
the Gaudin-Yang expression \eqref{eq:Delta_BCS} is not analytic for 
$g\rightarrow 0^-$ as a consequence of the divergence
of perturbation theory in the bulk even at vanishing interaction.

The few-body peculiarity of the functional dependence of $\Delta$
on $g$ is confirmed by the Bethe-ansatz result for two
paired fermions in a hard-wall trap of width $L$ \cite{Hao2006}.
In fact, the spin-excitation gap $\Delta$ for the hard-wall 
confinement potential shown in Fig.~\ref{figS1}(a)
(dashed curve) almost matches that for the harmonic trap (solid curve)
up to $g\sim -3$, provided that $L$ coincides with
the harmonic oscillator length 
$\ell=(\hbar/m\omega)^{1/2}$. At small $g$ the gap scales like
$\Delta \sim g^2$, which is clearly unrelated to the BCS-like functional
dependence of Eq.~\eqref{eq:Delta_BCS}.

\section{Cooper pairs}\label{s:pair}

\begin{figure}[htbp]
\setlength{\unitlength}{1 cm}
\begin{picture}(8.5,5.9)
\put(0.2,-0.2){\epsfig{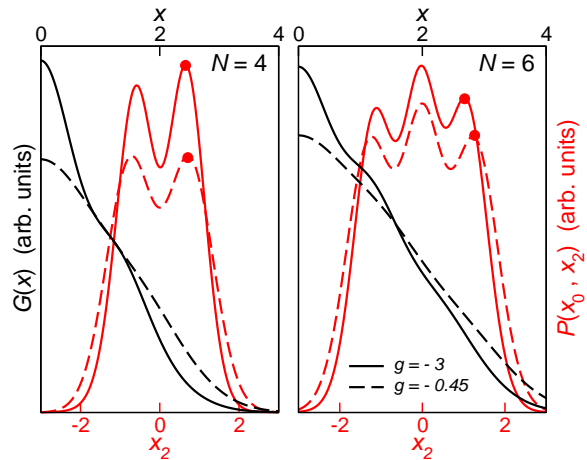}}
\end{picture}
\caption{(color online)
Conditional probability $P(x_0,x_2)$ vs $x_2$ (red [gray] curves,
right and bottom axes) and pair correlation function
$G(x)$ vs $x$ (black curves, left and top axes).
Left and right panels concern $N = 4$ and $N=6$,
respectively. Dashed (solid) lines correspond to $g = -0.45$ ($g=-3$).
Red [gray] dots locate the positions $x_0$ of spin-$\downarrow$ atoms.
The length unit is
$\ell = (\hbar/ m \omega)^{1/2}$.
}
\label{pair}
\end{figure}

To investigate pair formation
we evaluate the conditional probability $P(x_1,x_2)$ of finding 
one atom at position $x_2$ with spin $\sigma_2=\uparrow$ if another 
atom is fixed at $x_1$
with opposite spin $\sigma_1 = \downarrow$,
\begin{equation} \label{conditional_probability}
P(x_1,x_2) = A \sum_{i,j =1}^N 
\left< \delta(x_i-x_1)\delta_{\sigma_i,\downarrow} \delta(x_j-x_2)
\delta_{\sigma_j,\uparrow} \right>,
\end{equation}
where the quantum average $\left< \ldots \right>$  is taken
over the ED ground state
and $A$ is a normalization constant specified below. 
In Fig.~\ref{pair} we choose $x_1$ as the   
average radius $x_0=\left<\left|x_1\right|\right>$ (located
by the red [gray] dot) 
and plot
$P(x_1=x_0,x_2)$ versus $x_2$ (red [gray] curves).
At small interaction strength $g=-0.45$ (dashed red [gray] curves)
the conditional probability
is essentially independent from the fixed atom position 
$x_0$, thus replicating the spin-$\uparrow$ noninteracting one-body 
density $\sum_j\left<\delta(x_j-x_2)\right>$,
whose peaks are Friedel oscillations induced by the harmonic 
confinement \cite{Wang2012b,Soffing2011}.
For strong attraction, $g=-3$ (solid red [gray] curves), $P(x_0,x_2)$
rearranges its weight, 
exhibiting a clear shrinking of the lateral extension together
with a marked weight increase exactly at the position of the fixed atom. 
This suggests that the spin-$\uparrow$ atom at $x_2$ forms 
a bound Cooper pair with the spin-$\downarrow$ atom located at $x_0$.

To single out the internal structure of the Cooper pair
we average $P(x_1,x_2)$ over the center-of-mass coordinate 
$X=(x_1+x_2)/2$. The outcome is the pair correlation function 
\begin{equation}
G(x)
= \int_{-\infty}^{\infty}\!\!\!\! dX\, 
P(X+ x/2, X-x/2),
\label{eq:Gxdef}
\end{equation}
which is the probability of finding two atoms of opposite 
spins at the relative distance $x=x_1-x_2$. 
We choose the normalization constant $A$ of \eqref{conditional_probability}
to obtain $\int\! dx\, G(x)=1$. 
Figure \ref{pair} shows that $G(x)$ develops a dominant peak 
at the origin, whose height increases 
with the interaction---switching from $g=-0.45$
(dashed black curves) to $g=-3$ (solid black curves).
This tendency maximizes the spatial overlap of two atoms with 
opposite spins while suppressing the 
probability of finding them to separately wander in the trap.
Therefore $G(x)$ must be understood as the wave function square modulus 
of the Cooper pair in the frame of the relative distance $x$ between the 
two paired atoms, unrelated to the $X$-dependent distribution  
of pairs in the trap.
Note that the counterpart of $G(x)$ in the bulk
is the spatially-varying contribution to $P(x_1,x_2)$ explicited in 
\eqref{eq:PandG}, both quantities rapidly vanishing as $\left|x_1-x_2\right|
\rightarrow \infty$ [cf.~Fig.~\ref{Gx}(f)].

\begin{figure}[t]
\setlength{\unitlength}{1 cm}
\begin{picture}(8.5,5.7)
\put(0.2,-0.2){\epsfig{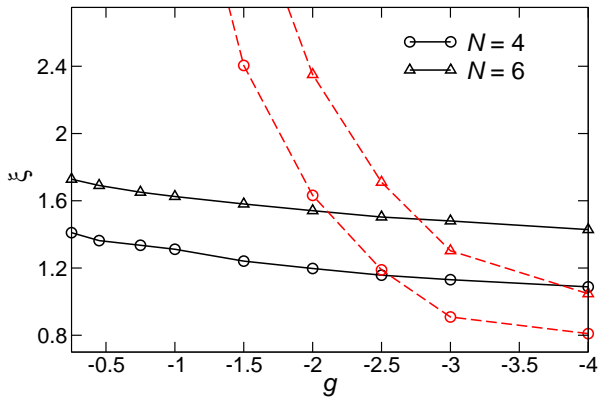}}
\end{picture}
\caption{
(color online)
Cooper pair size $\xi$ vs interaction strength $g$ for $N=4$ (circles) and
$N=6$ (triangles). 
Black (red [gray]) symbols with solid (dashed) lines point to the values of
$\xi_G$ ($\xi_c$).
We use $\hbar\omega$ as energy unit and 
$\ell = (\hbar/ m \omega)^{1/2}$ as length unit, hence
the dimensionless coupling constant is $g=g'/(\hbar\omega\ell)$.
The lines are guides to the eye.
}
\label{size}
\end{figure}

The Cooper pair size $\xi$ may be immediately obtained as the
quadratic displacement of $G(x)$,
\begin{equation} \label{pair_size}
\xi_G^2 = \int_{-\infty}^{\infty}\!\!\! dx\, x^2 G(x). 
\end{equation}
We see in Fig.~\ref{size} that $\xi_G$ depends only weakly on the atom number
(black symbols with solid lines)
and decreases with increasing attraction, as the pair motion becomes
more correlated.
For comparison, we also evaluate the pair size $\xi$ through the
BCS coherence length formula \cite{Brink2005}
(red [gray] symbols with dashed lines in Fig.~\ref{size}), 
which in standard units reads
\begin{equation} \label{pair_size_standard}
\xi_c= \frac{\hbar v_F}{2\Delta}.
\end{equation}
Here $v_F$ is the Fermi velocity obtained through the
equivalence $mv^2_F/2 = \Delta_1^*(N)$ and
$\Delta$ is taken from the ED excitation gap, $\Delta=E_{\text{exc}}(N)/2$.

In the noninteracting limit obviously
$\xi_c \rightarrow \infty$ (red [gray] symbols with dashed lines),
whereas $\xi_G$ (black symbols with solid lines)
tends to the natural limit fixed by the trap size.
However, for $\left|g\right|>2$ the two estimates become comparable,
pointing to a
BEC-like regime where the pair size $\xi$
is smaller than both trap size and interparticle spacing,
which could make correlated pair tunneling
observable \cite{Rontani2013b,Zuern2013}. This latter scenario contrasts with
the nuclear case, where the size of nucleon pairs is larger than the
system.

\section{Regime of strong attraction:
Pauli blocking in real space}\label{s:Pauli}

For strong inter-species attraction, the pair correlation function
$G(x)$ develops one (two) shoulder(s) for $N=4$ ($N=6$)
[black solid curve for $g=-3$ in the left (right) panel 
of Fig.~\ref{pair}]. 
This suggests that pairs arrange themselves in the trap
to minimize the residual pair-pair repulsion due to 
exchange forces acting between atoms with parallel spins,
as previously suggested in higher dimensions 
\cite{vonStecher2008,Rontani2009b,Bugnion2013c}.

\begin{figure}[htbp]
\setlength{\unitlength}{1 cm}
\begin{picture}(8.5,5.5)
\put(0.2,-0.2){\epsfig{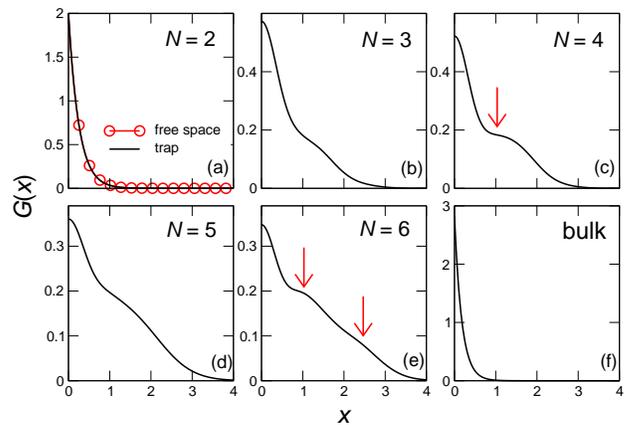}}
\end{picture}
\caption{(color online)
Pair correlation function 
$G(x)$ vs $x$ for (a) $N=2$ (b) $N=3$ (c) $N=4$ (d) $N=5$ (e) $N=6$
and (f) bulk at $g=-4$. Curves shown in panel a are analytical
whereas those in panels b-e are computed using 
a basis set made of 50 (panel b) and 25 (panels c-e) 
harmonic-oscillator levels, respectively. The bulk curve of panel f 
is the spatially-varying part of $G(x)$ obtained from the BCS wave function.
The length unit is
$\ell = (\hbar/ m \omega)^{1/2}$.
}
\label{Gx}
\end{figure}

The shoulders in the pair wave function become more evident
when dimers are strongly bound, as shown in Fig.~\ref{Gx}
for $g=-4$. 
The pair size $\xi$ is now
comparable to interparticle spacing, placing us on the BEC side
of the BCS-BEC  crossover
(cf.~Fig.~\ref{size}). 
This is also seen from the overlap between the 
wave function square modulus $G(x)$ in the trap 
[black line in Fig.~\ref{Gx}(a)] and 
in free space (circles) for a single pair, which 
is insensitive to the boundary as it is
squeezed by interaction. 
Consistently,
$\Delta=E_{\text{exc}}(N=2)/2 = 2.23$ 
in the trap matches the expectation of 
Gaudin-Yang model for $\gamma \rightarrow -\infty$, which is
half the binding energy  of a single dimer,
$\Delta = g^2/8=2$. 

Hovever, for more than one pair 
[Figs.~\ref{Gx}(b-e)] 
$G(x)$ qualitatively departs from the bulk prediction
shown in Fig.~\ref{Gx}(f)   
(derived in
Appendix \ref{a:Gbulk}). Whereas in 
the bulk $G(x)$ is a simple exponential, in the trap it 
displays $N/2
-1$ shoulders (with $N$ even), highlighted by arrows.
While the first shoulder already appears for $N=3$ [Fig.~\ref{Gx}(b)]
this feature is significantly strengthened for $N=4$ [Fig.~\ref{Gx}(c)],
as the available number of Cooper pairs increases in a
combinatorial fashion.
Besides, as a second shoulder becomes evident for $N=6$ [Fig.~\ref{Gx}(e)],
the first shoulder moves closer to the origin.
We attribute the overall behavior to Pauli blocking in real space,
since two atoms of like spin cannot occupy the same trap orbital in 
the relative frame. 
This structure, peculiar to the trap, 
may be measured by time-of-flight spectroscopy \cite{Bloch2008}.

\section{Conclusions}

In conclusion, we have studied a few 1D
Fermi atoms in the presence of attractive contact forces through
numerical exact diagonalization and found evidence of BCS-like paring.
Whereas the present 
experiments may be understood by treating the interaction
energy as a perturbation, we predict that non-trivial 
co-operative effects emerge at viable interaction strengths,
when the Cooper pair size compares with the trap size.

\begin{acknowledgments}
We thank Gerhard Z{\"u}rn, Stephanie Reimann,
Sven {\AA}berg, Frank Deuretzbacher, and Nikolaj Zinner for 
stimulating discussions. 
This work is supported by EU-FP7 Marie Curie initial training network 
INDEX, MIUR-PRIN2012 MEMO, and CINECA-ISCRA grant 
IscrC\_PAIR-1D.
\end{acknowledgments}

\appendix

\section{Bulk pair correlation function}\label{a:Gbulk}

In this Appendix we derive the expression of the
bulk pair correlation function $G(x)$
shown in Fig.~\ref{Gx}(f).
Throughout the Appendix we adopt standard units.

We introduce the BCS
wave function $\left|\Psi_{\text{BCS}}\right>$
as the bulk ground state, being
a standard variational ansatz in the whole
range of the BCS-BEC crossover \cite{Legget1980,Legget2006,Ketterle2008}.
In second quantization, $\left|\Psi_{\text{BCS}}\right>$ takes the form
\begin{equation}
\left|\Psi_{\text{BCS}}\right> = 
\prod_k \left(u_k + v_k\, \hat{c}^{\dagger}_{k\uparrow}
\hat{c}_{-k\downarrow}\right) \left| 0 \right>,
\label{Psi_BCS}
\end{equation}
where $\hat{c}^{\dagger}_{k\uparrow}$ is the fermionic operator that
acts on the vacuum $\left| 0 \right>$ creating an atom of spin
$\uparrow$ and momentum $k$. As usual, the BCS coherence factors
$u_k$ and $v_k$ occuring in \eqref{Psi_BCS} are defined as
\begin{displaymath}
u_k^2 = \frac{1}{2}\left( 1 + \frac{\xi_k}{E_k}\right),
\end{displaymath}
with $u_k^2 + v_k^2 = 1$. Here the quantity $\xi_k$ (not to be confused
with the pair size $\xi$) is
\begin{displaymath}
\xi_k = \varepsilon_k - \Delta_1,
\end{displaymath}
with $\varepsilon_k=\hbar^2k^2/2m$ being the single-particle energy
and $\Delta_1$ the bulk chemical potential,
and the quasiparticle energy $E_k$ is
\begin{displaymath}
E_k = \sqrt{ \xi_k^2 + \Delta^2 }.
\end{displaymath}
Depending on the value of the dimensionless coupling constant
$\gamma = (g\pi/\hbar)(m/8\varepsilon_F)^{1/2} $,
both chemical potentials $\Delta_1$ and pairing gap $\Delta$ should be
determined simultaneously \cite{Ketterle2008}.

The bulk conditional probability $P(x_1,x_2)$
of finding
one atom at position $x_2$ with spin $\uparrow$ if another
atom is fixed at $x_1$
with opposite spin $\downarrow$, analogous to the definition
(3) in the main text, is
\begin{equation}
P(x_1,x_2) = \left<\Psi_{\text{BCS}}\right| 
\hat{\Psi}^{\dagger}_{\uparrow}(x_2) \hat{\Psi}^{\dagger}_{\downarrow}(x_1)
\hat{\Psi}_{\downarrow}(x_1)\hat{\Psi}_{\uparrow}(x_2)
\left|\Psi_{\text{BCS}}\right>.
\label{eq:P_def}
\end{equation}
Here $\hat{\Psi}_{\sigma}(x)$ is the annihilation field operator
that destroys a fermion of spin $\sigma$ at position $x$:
\begin{equation}
\hat{\Psi}_{\sigma}(x) = \sum_k \frac{1}{\sqrt{L}}{\text{e}}^{ikx}
\hat{c}_{k\sigma},
\label{eq:psi_def}
\end{equation}
with $L$ being the system length.
Inserting this expansion
into \eqref{eq:P_def} and applying a standard manipulation,
which parallels Appendix D of Ref.~\onlinecite{BCS1957}, we obtain:
\begin{equation}
P(x_1,x_2) = \frac{N_{\uparrow}N_{\downarrow}}{L^2} + G(x_2 - x_1),
\label{eq:PandG}
\end{equation}
with $N_{\sigma}$ being the total number of atoms having spin $\sigma$.
The conditional probability \eqref{eq:PandG}
is the sum of a homogeneous background, $N_{\uparrow}N_{\downarrow}/L^2$,
due to uncorrelated atoms having opposite spins,
plus a spatially-dependent part, $G(x)$, which depends only on the
relative distance $x=x_2-x_1$.
Explicitly, one has:
\begin{equation}
G(x) = \frac{\Delta^2}{16\pi^2}
\int\!\! dk \int\!\! dk' \,
\frac{ {\text{e}}^{i(k-k')x} }
{E_k E_{k'} }.
\label{eq:Gx}
\end{equation}
This quantity may be regarded as the wave function square modulus
of the Cooper pair.

We now focus on the strongly attractive regime of Fig.~\ref{Gx}(f).
To proceed, we assume the pairing gap $\Delta$ to be the limit value
for $\gamma \rightarrow -\infty$, i.e. half the binding energy of a single
pair in free space, $\Delta = mg^2 /(8\hbar^2)$. In this limit we may neglect
the contributions of $\varepsilon_F$ and interpair
interactions to the chemical potential $\Delta_1$,
hence $\Delta_1 = - \Delta$ \cite{Ketterle2008}.
This allows us to expand
the quasiparticle energy keeping only the linear term in $\varepsilon_k$,
$E_k\approx \sqrt{\Delta_1^2 + \Delta^2}[1 - \Delta_1 \varepsilon_k
/(\Delta_1^2 + \Delta^2)] $.
Therefore, we may rewrite \eqref{eq:Gx} as $G(x)\propto I(x)^2$,
where
\begin{displaymath}
I(x) = \int\!\! dk \frac{ {\text{e}}^{ikx} }
{k^2 + k_{\text{BCS}}^2}
\end{displaymath}
is the Fourier transform providing a decaying exponential, with
\begin{equation}
k_{\text{BCS}}=\frac{m\left|g\right|}{\sqrt{2}\hbar^2}.
\end{equation}
The final form of the normalized BCS pair wave function in the limit
of strongly bound pairs, after dropping a prefactor, is:
\begin{equation}
G(x) = k_{\text{BCS}} {\text{e}}^{-2k_{\text{BCS}} \left|x\right|},
\label{eq:Gfinal}
\end{equation}
which is plotted in Fig.~\ref{Gx}(f).

It is interesting to compare \eqref{eq:Gfinal} with the wave function
square modulus $\left|\psi_{\text{dimer}}(x)\right|^2$
of a single pair in free space, which is shown in Fig.~\ref{Gx}(a) 
(circles):
\begin{equation}
\left|\psi_{\text{dimer}}(x)\right|^2 = k_{\text{dimer}} 
{\text{e}}^{-2k_{\text{dimer}} \left|x\right|}.
\label{eq:psifinal}
\end{equation}
This has the same form as \eqref{eq:Gfinal} except for the decay length
inverse,
\begin{displaymath}
k_{\text{dimer}}=\frac{m\left|g\right|}{2\hbar^2},
\end{displaymath}
which is a factor $\sqrt{2}$ smaller than $k_{\text{BCS}}$.
The shrinking of the pair size in the condensate is the effect of
the exchange forces affecting the
BCS many-body wave function \cite{Ketterle2008}.

%


%

\end{document}